\documentclass{nsart_eng}
\usepackage{cite}
\usepackage[dvips]{graphicx}
\usepackage{amsmath,amssymb}

\year{2012} \volume{0} \nomer{0} \firstpage{1}

\let\leq\leqslant
\let\ge\geqslant

\begin{document}

\setcounter{page}{1}
\newcommand{\re}[1]{(\ref{#1})}
\newcommand{\lab}[1]{\label{#1}}
\newcommand{\ci}[1]{\cite{#1}}
\renewcommand{\baselinestretch}{1.25}
\newcommand{\bfr}{\begin{flushright}}
\newcommand{\bfl}{\begin{flushleft}}
\newcommand{\efl}{\end{flushleft}}
\newcommand{\efr}{\end{flushright}}
\newcommand{\bc}{\begin{center}}
\newcommand{\ec}{\end{center}}
\newcommand{\be}{\begin{equation}}
\newcommand{\ee}{\end{equation}}
\newcommand{\bea}{\begin{eqnarray}}
\newcommand{\eea}{\end{eqnarray}}
\newcommand{\ba}{\begin{array}}
\newcommand{\ea}{\end{array}}
\newcommand{\edc}{\end{document}}
\newcommand{\ul}{\underline}
\newcommand{\ri}{\rightarrow\infty}
\newcommand{\li}{\leftarrow\infty}
\newcommand{\ra}{\rightarrow}
\newcommand{\la}{\leftarrow}
\newcommand{\ds}{\displaystyle}
\newcommand{\dsf}{\displaystyle\frac}
\newcommand{\dt}{\Delta{t}}
\newcommand{\il}{\int\limits}
\newcommand{\pal}{\partial}
\newcommand{\cal}{\mathcal}
\newcommand{\bone}{{\bf 1}}
\newcommand{\gComment}[1]{}
\renewcommand{\gComment}[1]{\textcolor{red}{Gerardo: #1}}

\title[{\it Intra pseudogap- and superconductivity pair spin...}]
{Intra pseudogap- and superconductivity-pair spin and charge fluctuations and underdome 
metal-insulator (fermion-boson)-crossover phenomena as keystones of cuprate physics}

\author[{\it B.~Abdullaev, D.\,B.~Abdullaev, C.\,-H.~Park, M.\,M. Musakhanov}]
{$^1$B.~Abdullaev, $^1$D.\,B.~Abdullaev, $^2$C.\,-H.~Park, $^3$M.\,M. Musakhanov}

\address{ 
$^1$Institute of Applied Physics, National University of Uzbekistan,
Tashkent 100174, Uzbekistan\\
$^2$Research Center for Dielectric and Advanced Matter Physics,
Department of Physics, Pusan National University, 30
Jangjeon-dong, Geumjeong-gu, Busan 609-735, Korea\\
$^3$National University of Uzbekistan,
Tashkent 100174, Uzbekistan}

\email{bakhodir.abdullaeff@yandex.ru,cpark@pusan.ac.kr,yousufmm@list.ru}


\begin{abstract}
The most intriguing observation of cuprate experiments is most likely the metal-insulator-crossover (MIC), seen in the underdome region of the temperature-doping phase diagram of copper-oxides under a strong magnetic field, when the superconductivity is suppressed. This MIC, which results in such phenomena as heat conductivity downturn, anomalous Lorentz ratio, nonlinear entropy, insulating ground state, nematicity- and stripe-phases and Fermi pockets, reveals the nonconventional dielectric property of the pseudogap-normal phase. Since conventional superconductivity appears from a conducting normal phase, the understanding of how superconductivity arises from an insulating state becomes a fundamental problem and thus the keystone for all of cuprate physics. Recently, in interpreting the physics of visualization in scanning tunneling microscopy (STM) real space nanoregions (NRs), which exhibit an energy gap, we have succeeded in understanding that the minimum size for these NRs provides pseudogap and superconductivity pairs, which are single bosons. In this work, we discuss the intra-particle magnetic spin and charge fluctuations of these bosons, observed recently in hidden magnetic order and STM experiments. We find that all the mentioned MIC phenomena can be obtained in the Coulomb single boson and single fermion two liquid model, which we recently developed, and the MIC is a crossover of sample percolating NRs of single fermions into those of single bosons.
\end{abstract}


\maketitle

\section{Introduction}

The origin of pseudogap (PG) and high-temperature superconductivity (HTS)
phases in copper oxides (cuprates) is one of the most puzzling and challenging problem
in condensed matter physics. Despite being almost three decades since their discovery,
intensive experimental and theoretical studies have yielded little clear understanding 
of these phases so far. The experimental studies of HTS and PG in cuprates have provided physicist 
by numerous interesting and fascinating materials with unconventional properties. Among 
the most puzzling and thus far most intriguing is the observation of the 
metal-insulator-crossover (MIC), seen in the underdome region of a temperature-doping 
phase diagram in the presence or absence of a strong external magnetic field~\ci{Takagi} 
\ci{Ando}. The MIC, detected after suppression of the HTS by a strong magnetic field, 
results in a number of different phenomena: heat conductivity downturn and anomalous Lorentz ratio~\ci{hill,proust}, 
nonlinear entropy~\ci{loram1,loram2}, insulating ground state~\ci{Takagi,Ando}, dynamic 
nematicity~\ci{Fujita} and static stripe phases~\ci{Vojta} and Fermi pockets~\ci{Vojta,Sebastian}. 
This reveals the highly unconventional dielectric property of the PG-normal phase of these 
superconductors. Since superconductivity appears in conventional superconductors from 
the conducting normal state only, the understanding of how HTS arises from an insulating state 
becomes a fundamental problem, and thus, the keystone for cuprate physics. This 
MIC also separates previously applied mechanisms and models for conventional superconductors 
from the consideration. However, the answer to the question "What quasiparticles do 
provide the PG and HTS phases?" still remains elusive. 

In Ref.~\ci{Abdullaev1} we demonstrated that these quasiparticles are PG and HTS pairs. We 
came to this conclusion when trying to interpret the physics visualized in STM real space PG 
and HTS nanoregions~\ci{Gomes,Pan}, which exhibited an energy gap.
In~\ci{Abdullaev1}, we showed that the minimally sized these nanoregions are real space pairs and furthermore, 
these pairs are single bosons. We also have a good qualitative understanding of all elements for the 
temperature-doping phase diagram of copper-oxides. We have shown that the precursor mechanism of the HTS 
is valid, when bulk superconductivity at the critical temperature $T_c$ or at the first critical 
doping $x_{c1}$ appears as a percolation phase transition for the spatial overlapping of 
separated PG pairs.

In this work we discuss the intra particle spin and charge fluctuations of PG and HTS pairs - single bosons, observed recently in hidden magnetic order~\ci{Fauque} and STM~\ci{Lawler} experiments. We find that all the above mentioned MIC phenomena might be obtained in the framework of the Coulomb single boson and single fermion two liquid model~\ci{Abdullaev1}, which naturally emerges from the analysis of the STM experimental data~\ci{Gomes,Pan} 
and thus draw a conclusion about the single boson nature of the visualized in the STM experiment pairs. 

Currently, we realize that the non-Fermi liquid property of these phenomena and thus of copper oxides is related 
to the mutual single boson and MIC physics. The insulating ground state of cuprates is a result of a gas of 
composed single bosons. According to the Bogoliubov approach for gas of charged bosons~\ci{Abdullaev5}, 
at high gas density, where this approach is valid, the ground state energy consists of components for the Bose-Einstein condensate and a gas of quasi-particles. 
At high magnetic fields or at lower levels with doping, to the first critical doping, 
for which all MIC phenomena are measured, the Bose-Einstein condensate vanishes and there, only a gas of quasi-particles exists, i.e., a gas of 2D plasmons of single charged bosons. However, the latter is insulating, therefore, the insulator is the whole ground state of copper oxides for underdome dopings from the first critical 
level ($x_{c1}$) up to the second critical level ($x_{c2}$). It is worth noting here that 
typically, the experiment detects the MIC up to the critical value of doping $x_c$, which, for
some cuprates, coincides with the optimal doping, at which the $T_c$ is maximum in the temperature-doping phase
diagrams, and for others with the second critical doping level $x_{c2}$. For the sake of simlicity, we assume 
that  $x_c$ coincides with $x_{c2}$. 

According to the phenomenological Coulomb single boson and single fermion two liquid model, fermions, 
which are responsible for the electric conductivity, emerge in the system at the first critical doping  
level $x_{c1}$. They are not active in the sense of contributing to the bulk electric conductivity in the range
between  $x_{c1}$ and $x_{c2}$. Only from the second critical doping level $x_{c2}$, beyond which the spatial 
percolation of NRs for fermions occurs, do these fermions contribute to penetration of charges in entire volume 
of a sample. From this analysis, we realize that the MIC is the fermion - boson crossover of sample 
percolating NRs of single fermions into that of single bosons. As it was shown below (see also Ref.~\ci{Abdullaev1}), NRs of single fermions start to percolate from $x_{c2}$ while those of single 
bosons do so from $x_{c1}$.

The insulating behaviour of fermions in the normal phase underdome region of doping results in the
insulating property of Fermi pockets, which is a generalization for Fermi pockets, seen in 
angle resolved photoemession spectroscopy (ARPES) experiment~\ci{Sebastian} (for details see 
Sec.~\ref{sec9}). The observed underdome doping evolution of these pockets qualitatively coincides 
with that of a gas of fermions in the Coulomb two liquid model. These insulating fermions also result in 
an insulating stripe phase (the smectic phase), seen in the PG phase of some copper oxides~\ci{Vojta}, 
when fermion charges, with their collective static electric field, deform the parent compound lattice, 
consisting of electrically polar atoms. 

It is interesting that the spatially intra rare charge density of each single boson (see the 
size of last one in the table of Sec.~\ref{sec6}) allows one to understand the nature of the 
intra-unit cell nematic order (the dynamic charge fluctuations) for the sample hole density 
and its evolution with doping, which was recently observed in the STM experiment~\ci{Lawler}. 
This is also the result of the strong ferrielectric crystal field of the parrent compound and 
the decrease of single bosons with doping.

The interesting hidden magnetic order experiment~\ci{Fauque} has revealed the existence in the
PG phase objects with zero total spin but with its fluctuations inside. Despite the authors' 
(Ref.~\ci{Fauque}) interpretation of the physical meaning of both spins, by suggesting either 
a pair of oppositely flowing 
intra-structural cell loop-currents or staggered spins in the same cell, the role of these objects 
in cuprate physics was not understood. One could assume that these objects are 
the PG and HTS pairs (for justification of this assumption see Sec. \ref{sec7} below), however, 
the two mentioned mechanisms for explaning the object spins were unable to provide the large 
experimental value for the pair energy gap which was observed for cuprates. This shortcoming was 
improved upon in our research~\ci{Abdullaev2,Abdullaev3}, which appeared in publication in the same year 
as the experiment described in~\ci{Fauque} and reproduced the experimental values of the $T_c$. In 
these papers, we have succeeded in understanding not only the nature of both object spins but also predict 
their evolution with doping. This evolution was successfully observed in further experiments on the 
hidden magnetic order.

In Sec. \ref{sec2} we demonstrate the rigorous proof that $2D$
fermions can bosonize. Then, in Sec. \ref{sec3}, the results for the
ground-state energy calculations of a charged anyon gas will be given.
We apply the difference between the ground state energies of
fermions and bosons to derive the single boson doping-temperature
phase diagram of cuprates in Sec. \ref{sec4}. In Sec. \ref{sec5}, it
will be demonstrated that this difference in the ground state
energies yields the microscopic origin of the phenomenological
Uemura relation. Sec. \ref{sec6} will be devoted to the charge and
percolation analysis of NRs on the base of experimental data given
in Refs.~\ci{Gomes} and~\ci{Pan}. This analysis provides the
interpretation of some elements of the phase diagram
doping-temperature in $Bi_2Sr_2CaCu_2O_{8+\delta}$ compound. 
In Sec.~\ref{sec7}, we will demonstrate that intra-structural cells,
for which the hidden magnetic order has been observed, are NRs, as 
displayed by the STM experiment~\ci{Gomes} and thus single bosons. 
We also show in this section that intra cell spin fluctuations 
and an electronic nematicity~\ci{Lawler} can naturally be 
understood within the single boson model. The MIC and insulating 
ground state, observed in a set of copper oxides, and their 
possible understanding within the Coulomb two liquid model is 
subject of Sec.~\ref{sec8}. The origin of Fermi pockets and 
stripe phases, seen experimentally in some cuprates, is discussed in 
Sec.~\ref{sec9} for our model. The answer for the interesting question: 
"Why is the ground state of YBCO copper oxides in a strong magnetic field an oscillating Fermi 
liquid, while for $Bi_2Sr_2CaCu_2O_{8+\delta}$ cuprate, it is 
insulating?" can be found in the Sec.~\ref{sec10}. Sec.~\ref{sec11} will 
describe the origin of the non-Fermi liquid heat conductivity and the entropy 
of copper-oxides. We summarize and conclude our paper in Sec. \ref{sec12}.

\section{Real Bosonization of $2D$ Fermions}
\label{sec2}

The $2D$ topology allows
fractional statistics \ci{lei}, characterized by a continuous
parameter $\nu$, having values between 0 (for bosons) and 1 (for
fermions). The particles with $0<\nu<1$ are generically called
anyons \ci{wil2}. One can apply the last criterion to investigate the
properties of the $a - b$ planes of $CuO_2$ atoms, which play a 
dominant role in the determination of cuprate physics.

In a manner similar to Ref.~\ci{Abdullaev1}, in this section we briefly
outline the rigorous derivation of the real bosonization of 2$D$
fermions. This can be achieved by exact cancellation of terms in the
Hamiltonian arising from fermion (anyon) statistics and a
Zeeman interaction of spins $\hbar /2$ of particles with statistical
magnetic field \ci{Dunne} produced by vector potential of
anyons.

Let us consider the Hamiltonian 
\bea \ba{r} \hat
H=\dsf{1}{2M}\ds\sum_{k=1}^N\left[\left(\vec p_k+\vec A_{\nu}(\vec
r_k)\right)^2+ M^2\omega_0^2 |\vec{r_k}|^2 \right] \\
+ \dsf{1}{2}\ds\sum_{k=1}^N\left[V(\vec r_k) +
\ds\sum_{j\not=k}^N\dsf{e^2} {|\vec r_{kj}|}\right] \ \lab{gsetup1}
\ea 
\eea 
of the gas of $N$ anyons with mass $M$ and charge $e$,
confined in a $2D$ parabolic well, interacting through Coulomb
repulsion potential in the presence of uniform positive
background~\ci{Laughlin} $ V(\vec r_k)$. Here, $\vec r_k$ and $\vec
p_k$ represent the position and momentum operators of the $k$th
anyon in 2$D$ space dimension, 
\be \vec A_{\nu }(\vec
r_k)=\hbar\nu\ds\sum_{j\not=k}^N\dsf{\vec e_z \times\vec r_{kj}}
{|\vec r_{kj}|^2} \lab{gsetup2} 
\ee 
is the anyon gauge vector
potential \ci{wu}, $\vec r_{kj}=\vec r_k-\vec r_j$, and $\vec e_z$
is the unit vector normal to the 2$D$ plane. In the expression for
$\vec A_{\nu }(\vec r_k)$ and hereafter we assume that $0\leq \nu
\leq 1$.

In the bosonic representation of anyons we take the system wave
function in the form~\ci{Comtet}: 
\be \Psi(\vec
R)=\prod_{i\not=j}r_{ij}^{\nu}\Psi_T(\vec R). \lab{gsetup7} 
\ee 
Here
$\vec R = \{\vec r_1....\vec r_N\}$ is the configuration space of
the $N$ anyons. The product in the right hand side of this equation
is the Jastrow-type wave function. It describes the short distance
correlations between two particles due to anyonic (fermionic)
statistical interaction.

Let us consider first the term in the Hamiltonian $\hat H$,
Eq.~\re{gsetup1}, containing the anyon vector potential $\vec A_{\nu
}(\vec r_k)$. Substituting $\Psi(\vec R)$, Eq.~\re{gsetup7}, into the
Schr\"odinger equation with this Hamiltonian, we obtain an equation
$\widetilde{\hat H}\Psi_T(\vec R)=E\Psi_T(\vec R)$ with the novel
Hamiltonian $\widetilde{\hat H}=\widetilde{\hat H}_1+\widetilde{\hat
H}_2$, where 
\be 
\widetilde{\hat
H}_1=\ds\sum_{k=1}^N\left(\dsf{-\hbar^2\Delta_k}{2M}-
\dsf{\hbar^2\nu}{M}\ds\sum_{j\not=k}\dsf{{\vec r}_{kj}\cdot {\vec
\nabla}_k}{|\vec r_{kj}|^2}\right) \lab{gsetup7b} 
\ee
and 
\be \widetilde{\hat
H}_2=-i\dsf{\hbar}{M}\ds\sum_{k=1}^N\left(\vec A_{\nu }(\vec
r_k)\cdot {\vec \nabla}_k+\nu \ds\sum_{j\not=k}\dsf{\vec A_{\nu
}(\vec r_k)\cdot {\vec r}_{kj}}{|\vec r_{kj}|^2}\right).
\lab{gsetup7c} 
\ee

As shown in Ref.~\ci{Comtet}, the $\nu$ interaction Hamiltonian in
$\widetilde{\hat H}_1$, i.e., the second its term, is equivalent to
a sum of two-body potentials (accuracy of this term discussed in~\ci{Abdullaev1})
\be 
\dsf{\pi\hbar^2 \nu
}{M}\ds\sum_{j\not =k} \delta ^{(2)}( \vec r_k- \vec r_j) \ .
\lab{gsetup7aa} 
\ee 
Therefore, the Hamiltonian $\widetilde{\hat
H}_1$ now reads
\be 
\widetilde{\hat
H}_1=\ds\sum_{k=1}^N\left(\dsf{-\hbar^2\Delta_k}{2M}+
\dsf{\pi\hbar^2 \nu }{M}\ds\sum_{j\not =k} \delta ^{(2)}( \vec r_k-
\vec r_j)\right). \lab{gsetup7ab} 
\ee

Now we demonstrate the real bosonization of 2$D$ fermions for the
example of anyons in a parabolic well. To do this, we consider the
Zeeman interaction term 
\be 
\dsf{\hbar }{M}\ds\sum_{k=1}^N {\hat
{\vec s}} \cdot \vec b_k \ \lab{gsetup8} 
\ee 
of spins with the
statistical magnetic field \ci{Dunne} 
\be 
\vec
b_k = -2\pi \hbar \nu \vec e_z \ds\sum_{j\not =k} \delta ^{(2)}(
\vec r_k- \vec r_j) \ , \lab{gsetup9} 
\ee 
which can be derived if
one calculates $\vec b_k = \vec \nabla \times \vec A_{\nu }(\vec
r_k)$ by using Eq.~\re{gsetup2}.

For $s_z=\hbar /2$, and using the expression, Eq.~\re{gsetup9}, for
$\vec b_k$ one obtains the following for the
Zeeman term:
\be 
\dsf{\hbar }{M}\ds\sum_{k=1}^N {\hat
{\vec s}} \cdot \vec b_k=
 -\pi  \nu \dsf{\hbar^2}{M} \ds\sum_{k(j\not =k)}
\delta ^{(2)}( \vec r_k- \vec r_j) \ . \lab{gsetup10} 
\ee
It is easy to see that the contribution to the energy from  the Hamiltonian
$\widetilde{\hat H}_2$ is zero. On the other hand, this expression for the
Zeeman term, added to Eq.~\re{gsetup7ab}, exactly cancels the second
term of $\widetilde{\hat H}_1$, which is responsible for 
fermion (for $\nu=1$)  and anyon statistics. Since the energy
of bosons is lower than that of fermions and anyons, there appears a
coupling of spin with statistical magnetic field for every particle
or bosonization of 2$D$ fermions and anyons. From this, one can
conclude, if anyon concept is correct for the description of any
$2D$ quantum system, its ground state should be bosonic with
$\nu=0$, while its excited state should be fermionic ($\nu=1$) or
anyonic ($0<\nu<1$) depending upon the fixed value for $\nu$. 

\section{The ground-state energy of charged anyon gas} \label{sec3}

In Ref.~\ci{Abdullaev5}, we derived an approximate analytic
formula for the ground-state energy of a charged anyon gas. Our
approach was based on the harmonically confined two-dimensional
(2$D$) Coulomb anyon gas and a regularization procedure for
vanishing confinement. To take into account  the fractional
statistics and Coulomb interaction, we introduced a function, which
depends on both the statistic and density parameters ($\nu$ and
$r_s$, respectively). We determined this function by fitting our 
energy to the ground state energies of the classical electron crystal at very
large $r_s$ (the 2$D$ Wigner crystal), and to the Hartree-Fock (HF)
energy of the spin-polarized 2$D$ electron gas, and the dense 2$D$
Coulomb-Bose gas at very small $r_s$. The latter was calculated by
use of the Bogoliubov approximation. When applied to the boson system
($\nu=0$), our results were very close to those obtained recently from Monte
Carlo (MC) calculations. For spin-polarized electron systems
($\nu=1$), our comparison led to a critical judgment concerning the
density range, to which the HF approximation and MC simulations
apply. 

We have found in Ref.~\ci{Abdullaev5} the expression for the
ground-state energy per particle (in Ry units) in the form 
\be 
{\cal E}_0(\nu, r_s)\approx
\dsf{2f(\nu,r_s)}{r_s^2}\left[\dsf{\nu}{2K_X^2}+\dsf{K_X^2}{2}-
\dsf{K}{K_X} \right] \, . \lab{gsqa11} \ee Here \bea \ba{r}
K_X=(K_A+K_B)^{1/2}\\[1mm]
 +[-(K_A+K_B)+2(K_A^2-K_A
K_B+K_B^2)^{1/2}]^{1/2} \, , \lab{gsqa6a} \ea \eea and \bea \ba{l}
K_A=\left[K^2/128+\left((\nu/12)^3+(K^2/128)^2
\right)^{1/2} \right]^{1/3},\\
K_B=\left[K^2/128-\left((\nu/12)^3+(K^2/128)^2 \right)^{1/2}
\right]^{1/3}, \ \lab{gsqa9} 
\ea 
\eea 
with $K=c_{WC}r_s/f^{1/2}(\nu,r_s)$.

For the Bose gas ($\nu = 0$), we obtained from Eq. ~\re{gsqa11}  
\be
{\cal E}_0(0, r_s)=-\dsf{c_{WC}^{2/3} f^{2/3}(0,r_s)}{r_s^{4/3}} \, .
\lab{gsqa12} 
\ee 
We found for small $r_s$ $f(0,r_s)\approx
c_{BG}^{3/2}r_s/c_{WC}$, which fitted the ground-state energy of
Coulomb Bose gas calculated in Ref.~\ci{Abdullaev5} using the
Bogoliubov approximation with $c_{BG}=1.29355$. For large $r_s$, the
ground-state energy does not depend on statistics and equals the
energy of the classical 2$D$ Wigner crystal \ci{bm},
$E_{WC}=-2.2122/r_s$. This matches with Eq.~\re{gsqa12} if at low
densities $f(0,r_s)\approx r_s^{1/2}$ with $c_{WC}^{2/3}=2.2122$.
For large $r_s$ we obtained 
\be 
{\cal E}_0(\nu, r_s \rightarrow
\infty ) =\dsf{c_{WC}^{2/3} f^{2/3}(\nu,r_s)} {r_s^{4/3}}\left(-1+
\dsf{7\nu f^{2/3}(\nu,r_s)}{3c_{WC}^{4/3}r_s^{4/3}}\right) \, .
\lab{gsqa15} 
\ee

For arbitrary $r_s$, the interpolating functional form is:
\bea 
\ba{r}
f(\nu,r_s)\approx \nu^{1/2}c_0(r_s)e^{-5r_s}\\[1mm]
+\dsf{c_{BG}^{3/2}r_s/c_{WC}}
{1+c_1(r_s)c_{BG}^{3/2}r_s^{1/2}/c_{WC}}
 +\dsf{0.2
c_1(r_s)r_s^2 Ln(r_s)}{1+ r_s^2} \,  \lab{gsqa17} 
\ea 
\eea 
with
$c_0(r_s)=1+6.9943r_s+22.4717r_s^2$ and $c_1(r_s)=1-e^{-r_s}$
satisfying all constraints for $f(\nu,r_s)$ function (see
Ref.~\ci{Abdullaev5}) and, in addition, yields in the fermion case
($\nu=1$) for the ground-state energy per particle the HF
result~\ci{rajagopal} 
\be 
E_{HF}=\dsf{2}{r_s^2}- \dsf{16}{3\pi r_s}
\, . \lab{gsqa18} 
\ee

In Fig. \ref{fig1} we show results for the ground-state energy per particle
on the large scale $1.0\leq r_s \leq15.0$. The upper four curves
refer in descending order to the fermion cases ($\nu =1$): HF
energies for spin-polarized electrons from Ref. \ci{rajagopal} (open
triangles), interpolated by Pad$\grave e$ approximant MC data from
Ref.~\ci{tanatar} for spin-polarized electrons (crosses), and our
results from Eq. ~\re{gsqa11} (on the given scale identical with
those of Eq. ~\re{gsqa15}) for spin-polarized electrons (closed
circles). The lower two curves are for charged bosons ($\nu =0$) and
result from MC calculations of Ref. \ci{DePalo} (closed triangles)
and from our Eq. \re{gsqa12} (open squares). By star symbols, we
indicated the MC data \ci{attac} (without interpolation) obtained
for some particular $r_s$ values.

\begin{figure}
\begin{center}
\includegraphics[angle=270,width=9.5cm,scale=1.0]{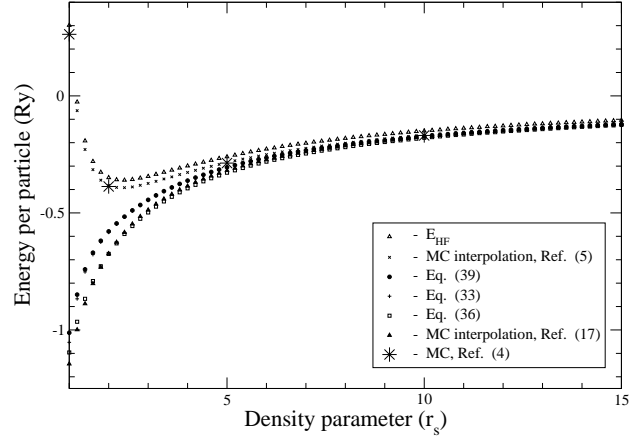}
\end{center}
\caption{Ground-state energies per particle vs density parameter
$r_s$ values ranging from $1.0\leq r_s \leq 15.0$ from top to bottom: for
fermions ($\nu = 1$) HF approximation (Eq. ~\protect\re{gsqa18} and
Ref. ~\protect\ci{rajagopal}, open triangles), MC interpolation data
(from Ref. ~\protect\ci{tanatar} (Ref.[5]
in~\protect\ci{Abdullaev5})), (crosses), and present results from
Eq. ~\protect\re{gsqa15}  (Eq.(39) in~\protect\ci{Abdullaev5})
(closed circles) and Eq. ~\protect\re{gsqa11} (Eq.(33)
in~\protect\ci{Abdullaev5}) (plus signs), and for bosons ($\nu = 0$)
present results from Eq. ~\protect\re{gsqa12} (Eq.(36)
in~\protect\ci{Abdullaev5}) (open squares) and MC data from Ref.
~\protect\ci{DePalo} (Ref.[17] in ~\protect\ci{Abdullaev5}) (closed
triangles). MC data of Ref. ~\protect\ci{attac} (Ref.[4] in
~\protect\ci{Abdullaev5}) for some particular values of $r_s$ are
indicated by star symbols.} \lab{fig1}
\end{figure}

\section{Single Boson doping-temperature phase diagram} 
\label{sec4}

Following to Refs.~\ci{Abdullaev2,Abdullaev3}, one can assume that
coupled to anyon magnetic field spins of fermions are fluctuating.
Therefore, bosons with effective spins might appear as Fermi
particles. However, fermions with different spins are independent
\ci{landaus}. Thus, the spins of bosons interact with each other and
do not interact with spins of another fermions if they exist in the
system. We introduce a correlation length, inside of which
spins of bosons interact with each other. For temperature $T=0$ we
denote it by $\xi_0$. The increase of fluctuations destroys the
coupling, and bosons become the anyons or fermions. This occurs when
the gain in the energy due to fluctuations of spins of bosons is
equal to energy difference between the anyon (or Fermi) and Bose
ground states.

For the interaction of boson spins, we introduce the following form: 
\be
e^{-r_0/\xi_0 }  \ds\sum_{k=1}^N {\hat {\vec s}}_{k+\delta} \cdot
{\hat {\vec s}}_k \ . \lab{gstup13} 
\ee 
Here, a
factor $e^{-r_0/\xi_0 }$ was introduced with $r_0$ is being the mean distance
between particles. For screening by magnetic field spins, $\xi_0$ is
assumed to be phenomenological and taken from the experiment.

We establish the explicit form of Eq. ~\re{gstup13}. The growth of
boson spin fluctuations should cancel term, Eq. \re{gsetup8}, in the
Hamiltonian. Therefore, for a dense Bose gas ($r_0<\xi_0$), the following 
should hold: ${\hat {\vec s}}_{k+\delta}=-\hbar \vec b_k/M$.

The Hamiltonian of a bosonized infinite anyon Coulomb gas with
spin interaction has a form: 
\bea 
\ba{r} 
\hat H=\dsf{1}{2M}\ds\sum_{k=1}^N\left[\left(\vec p_k+\vec A_{\nu}(\vec
r_k)\right)^2+ M V(\vec r_k)\right] \\
+\dsf{1}{2}\ds\sum_{k,j\not=k}^N\dsf{e^2} {|\vec r_{kj}|}+
\dsf{\hbar (1-e^{-r_0/\xi_0 }) }{M}\ds\sum_{k=1}^N {\hat {\vec s}}
\cdot \vec b_k  \ . \lab{gstup14}  
\ea 
\eea

For an anyon Coulomb gas with density parameter $r_s>2$, where $r_s$ is
$r_0$ in Bohr radius $a_B$ units, the approximate ground state
energy per particle is expressed by Eq.~\re{gsqa15}. In our
treatment we consider the bosonized fermions with $\nu=1$. To become
fermions, bosons should overcome the energy difference: 
\be
\Delta_0^B = \dsf{7 (1- e^{-r_s/\xi_0 }) f^{4/3}(0,r_s)}
{3c_{WC}^{2/3}r_s^{8/3}} \ , \lab{gstup17} 
\ee 
i.e., the 
superconductivity gap. Our approach in~\ci{Abdullaev5} corresponds to
spinless or fully spin-polarized fermions. Our system is a normal, i.e.,
non-spin-polarized electron liquid. However, for intermediate values of 
$r_s$, in which we have interest, according to Ceperley MC data~\ci{tanatar},
there is a tiny difference of ground state energies for spin-polarized and
non-spin-polarized electrons. Therefore, we can use ground-state energy for
spin-polarized fermions.

\begin{figure}
\begin{center}
\includegraphics[angle=0,width=9.5cm,scale=1.0]{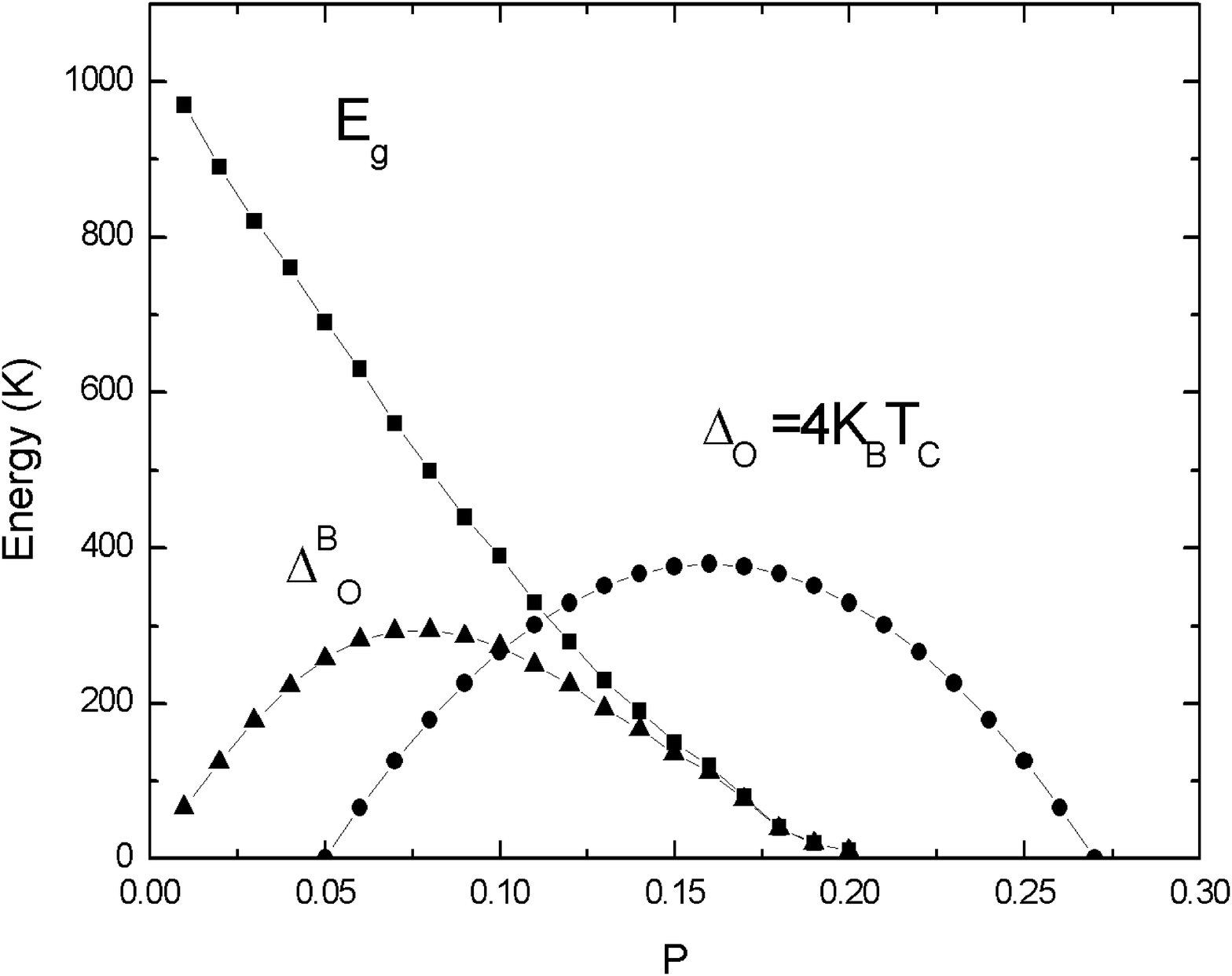}
\end{center}
\caption{ The experimental PG $E_g$, superconductivity gap
$\Delta_0=4 K_B \ T_c$ (experiment for hole doped $Bi$ \ - 2212
compound), and calculated from formula Eq. ~\protect \re{gstup17}
one for bosons $\Delta_0^B$ energies in Kelvin temperature (K) units
as function of concentration of holes $p$. } \lab{fig2}
\end{figure}

The Fig. \ref{fig2} displays  the PG boundary energy $E_g$ (Fig. 11 from
paper \ci{tallon}), superconductivity gap energy $\Delta_0=4 K_B \
T_c$, which was evaluated using the empirical formula
$T_c=T_{c,max}[1-82.6(p-0.16)^2]$ with $T_{c,max}=95 \ K$ for $Bi $
\ -2212 \ ($Bi_2 Sr_2 Ca Cu_2 O_{8+\delta}$) compound, and energy
gap calculated from Eq. ~\re{gstup17} as function of doping $p$. As
seen from this figure, our $\Delta_0^B$ has the same magnitude as
experimental gap, but is qualitatively different from generally
accepted "dome" like doping-temperature phase diagram. However, it
is  in accordance with Fig. 10 of paper \ci{tallon} of Tallon and
Loram and their conclusion that  PG energy $E_g$ up to $p_c\approx
0.19$ separates Bose-Einstein condensate into regions, where density
of Cooper pairs is small and big (weak and strong
superconductivity).

\begin{figure}
\begin{center}
\includegraphics[angle=0,width=9.5cm,scale=1.0]{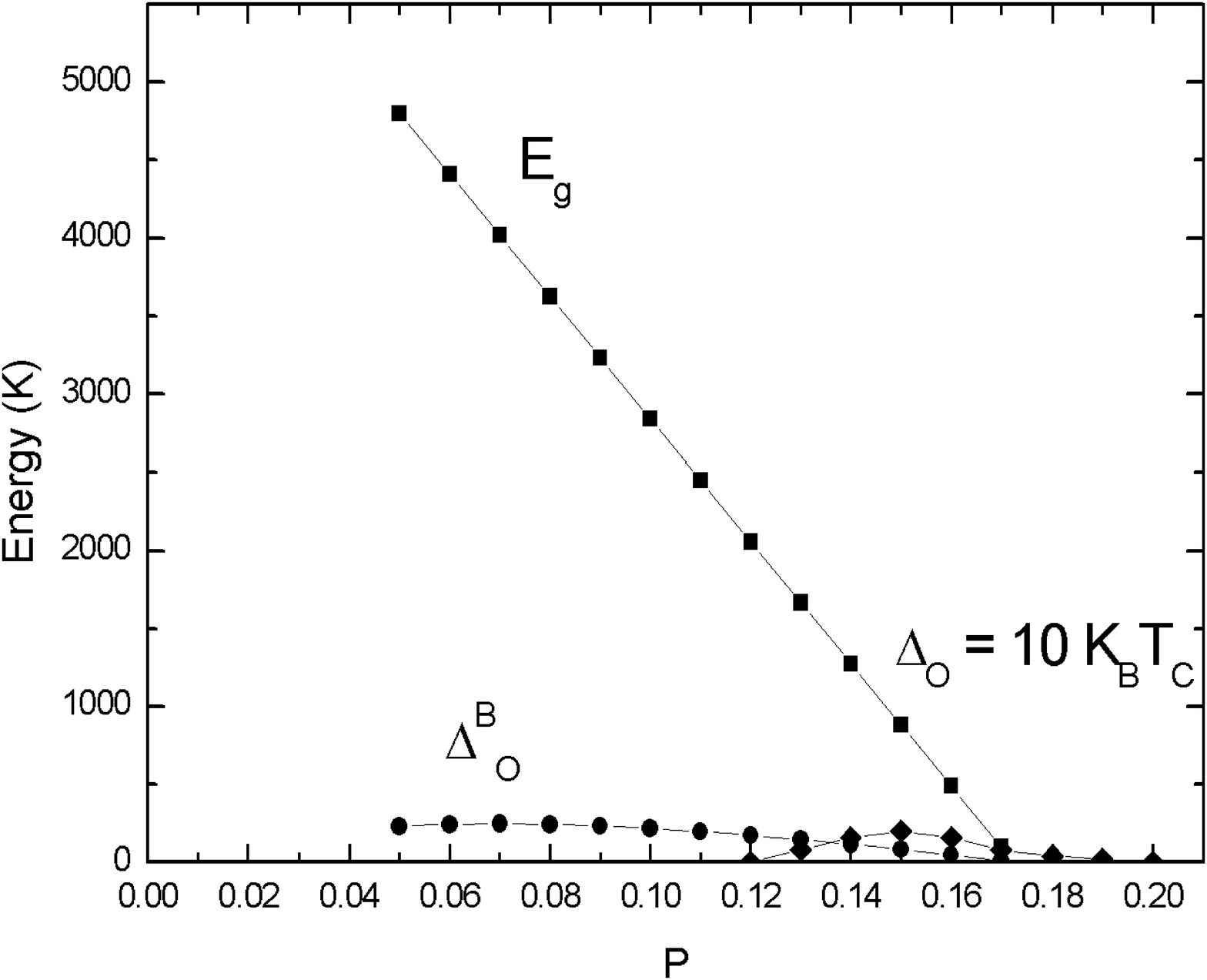}
\end{center}
\caption{ The experimental PG $E_g$, superconductivity gap
$\Delta_0=10 K_B \ T_c$ (experiments for electron doped NCCO and
PCCO compounds), and calculated from formula Eq. ~\protect
\re{gstup17} one for bosons $\Delta_0^B$ energies in Kelvin
temperature (K) units as function of concentration of electrons $p$.
} \lab{fig3}
\end{figure}

For phase diagram data of electron doped cuprates we use Ref.
\ci{onose} for NCCO ($Nd_{2-x} Ce_x Cu O_4$) and Ref. \ci{zimmers}
for PCCO ($Pr_{2-x} Ce_x Cu O_4$). It was shown there that $E_g/(k_B
T^*)\approx 10$ for NCCO and $E_g/(k_B T^*)\approx 11$ for PCCO,
therefore, we assume $E_g/(k_B T^*)\approx 10$ for both materials.
For experimental superconductivity gap we also assume $\Delta_0/(k_B
T_c)\approx 10$. Fig. \ref{fig3} shows the doping $p$ dependence of
experimental $E_g$, $\Delta_0 = 10k_B T_c$ and $\Delta_0^B$
calculated from Eq. ~\re{gstup17} by using the above spacing
constants of $a$ and $b$ for elementary structural cell. Comparing
with Fig. 2, we see the same qualitative and quantitative result.
More obvious is extension of our $\Delta_0^B$ to small values of
$p$, while experimental $\Delta_0$ starts with $p=0.13$. However,
absolute values of both superconductivity gaps of hole and electron
doped materials are nearly equal.

\section{Origin of Uemura Relation}
\label{sec5}

In this section, following to Ref.~\ci{Abdullaev1}, we describe the
single boson origin of the phenomenological Uemura relation for $2D$
superconductors.

Currently, it is widely accepted (see Ref.~\ci{zaanen}) that the Uemura
relation (UR), i.e., the linear dependence of $T_c$ on the concentration of the
charge carriers, originally observed in Refs.~\ci{uemura1} and
\ci{uemura2} for underdoped cuprate, bismuthate, organic,
Chevrel-phase and heavy-fermion superconductors, also survives for the
extended class of other superconductors and has a fundamental
universal character. An
experiment clearly relates the UR with $2D$ geometry of samples.
Motivated by this observation, in Ref.~\ci{Abdullaev1}, we
investigated the possible role of the fermion bosonization, which is
a result of the topology of $2D$, to the origin of UR.

The experimental doping dependence of $r_0$, mean distance between
two holes, can be approximated by the relationship $r_0\approx
a/x^{1/2}$ (see Fig. 34 in Ref.~\ci{Kastner}), where $a$ is a
lattice constant in an elementary structural plaquette for the
$CuO_2$ $a - b$ plane of a copper oxide. Since in~\ci{Abdullaev1}
the doping value denotes by the variable $x$, we keep this notation
in this and another section below, which are written on the basis of 
Ref.~\ci{Abdullaev1}. This relationship is derived in
Ref.~\ci{Kastner} for $La_{2-x}Sr_xCuO_4$ compound with $a\approx
3.8 \AA$. This lattice constant $a$ is the nearly same for other
copper oxide compounds, thus it is also valid for investigating here
compound $Bi_2Sr_2CaCu_2O_{8+\delta}$. It is worth noting that
$b\approx a$ for the lattice constant $b$ of the same structural
plaquette.

Applying the relationship $r_0\approx a/x^{1/2}$, where $a\approx
3.8 \AA$, we estimate values of $r_0$, expressed in Bohr
radius $a_B$ unit ($r_s=r_0/a_B$), corresponding to the doping interval
$x_{c1}\leq x\leq  x_{c2}$, where $x_{c1}$ and $x_{c2}$ are the
first and second critical dopings in the doping-temperature phase
diagram. When doing so, one sees that $13.12\leq r_s\leq 32.14$. For this interval of
$r_s$, we have obtained the expression, Eq. ~\re{gsqa15}, for the
ground state energy per particle of the Coulomb-interacting anyon
gas. It is expressed in $Ry$ (Rydberg)  energy unit and for large
$r_s$ equals the energy of the classical 2$D$ Wigner crystal
\ci{bm}, $E_{WC}=-c_{WC}^{2/3}/r_s$ with $c_{WC}^{2/3}=2.2122$.

Taking into account from the previous section that the excited state
of the $2D$ system is fermionic and the ground state is bosonic, one
can write the explicit expression for an energy gap between these
two states in the following manner:
\be 
\Delta(r_s)={\cal E}(\nu=1,r_s)-{\cal
E}(\nu=0,r_s)=\dsf{7 E_{WC}^2}{3c_{WC}^2} \ . \lab{gsetp12} 
\ee 
The
meaning of this expression in that to become a fermion the boson
should gain the energy $\Delta(r_s)$. Substituting in
Eq.~\re{gsetp12} the expression for $E_{WC}$ and introducing the
$2D$ density $n=1/(\pi r_0^2)$ one derives 
\be 
\Delta(n)=\dsf{7 \pi
n a_B^2}{3c_{WC}^{2/3}} \ . \lab{gsetp13} 
\ee 
Since the critical
temperature $T_c$ is proportional to $\Delta(n)$, one can conclude
that the $2D$ topology driven bosonization of fermions may explain
the UR for a variety superconductors, whose physics is quasi - two
dimensional.

\section{Experiment implied single boson elements of the doping-temperature phase diagram}
\label{sec6}

Recently, Gomes {\it et} {\it al.}~\ci{Gomes} have visualized the
gap formation in NRs above the critical temperature $T_c$ in the
HTS $Bi_2Sr_2CaCu_2O_{8+\delta}$ cuprate. It has been
found that, as the temperature lowers, the NRs expanded in the bulk
superconducting state consisted of inhomogeneities. The fact that
the size of the inhomogeneity~\ci{Pan} is close to the minimum size
of the NR~\ci{Gomes} leads to a conclusion that the HTS
phase is a result of these overlapped NRs. In the present section we
reproduce the main results of Ref.~\ci{Abdullaev1}, where the charge 
and percolation regime analysis of NRs was performed and
shown that at the first critical doping $x_{c1}$, when the
superconductivity initiates, each NR carries the positive electric
charge of one (in units of electron charge), thus we attributed the NR to
a single hole boson, and the percolation lines connecting these
bosons emerged. At the second critical doping $x_{c2}$, when the
superconductivity disappears, our analysis demonstrated that the
charge of each NR equals two. The origin of $x_{c2}$ can be
understood by introducing additional normal phase hole fermions in
NRs, whose concentration appearing above $x_{c1}$ increases smoothly
with the doping and breaks the percolation lines of bosons at
$x_{c2}$. The latter resulted in the disappearance of the bulk bosonic
properties of the PG region, which explained the upper bound for
existence of vortices in Nernst effect~\ci{Wang}. Since~\ci{Gomes}
demonstrated the absence of NRs at the PG boundary, one can
conclude that along this boundary, as well as in $x_{c2}$, all
bosons disappear.

The authors of Ref.~\ci{Gomes} have visualized the NRs in the PG
region of $Bi_2Sr_2CaCu_2O_{8+\delta}$ compound at fixed hole
dopings $x=0.12,0.14,0.16,0.19,0.22$. It has been determined that
for $x=0.16$ and $x=0.22$ the minimum size of the NRs is
$\xi_{coh}\approx 1-3$ nm. The estimated minimum size of NRs,
$\xi_{coh}$, is about 1.3 nm in the superconducting phase~\ci{Pan}
($T_c=84 K$). Another notable result obtained in Ref. \ci{Pan} was
the observation of spatial localization of the dopped charges. The
charges were localized in the same area as NRs \ci{Pan} with the same
coherence length $\xi_{coh}$.

The principal part of our analysis in Ref.~\ci{Abdullaev1} has been
the doping $x$ dependence of the NR charge $(\xi_{coh}/r_0)^2$. We
started with the case of zero temperature. The parameter
$\xi_{coh}/r_0$ contained essential information for our
consideration. The factor $(\xi_{coh}/r_0)^2$ reduces to the
expression $x(\xi_{coh}/a)2$ which has a simple physical meaning: it
is a total electric charge of $(\xi_{coh}/a)^2$ number of plaquettes,
each having a charge $x$. Conversely, the parameter
$\xi_{coh}/r_0$ describes the average spatial overlapping degree of
two or more holes by one NR. If $\xi_{coh}/r_0>1$ then all NRs will
be in close contact with each other, thus providing the bulk
superconductivity in the percolation regime.

In the Table I, we outlined the doping $x$ dependencies for the
function $(\xi_{coh}/r_0)^2$ for fixed experimental values
$\xi_{coh}=10\AA$ (the minimal size of the NR) and
$\xi_{coh}\approx13\AA$ taken from Ref.~\ci{Gomes} and
Ref.~\ci{Pan}, respectively, and for the function $\xi_{coh}$ which
fits $(\xi_{coh}/r_0)^2$ to $(10\AA/r_0)^2$ at $x=0.28$ and for
$x=0.05$ provides $(\xi_{coh}/r_0)^2\approx 1.0$. Numerical values
of the $\xi_{coh}/r_0$ are also shown in the table.

\begin{table}[tb]
\begin {center}
\begin{tabular}{|c|c|c|c|c|c|c|} \hline
    $x$   & $(10 \AA/r_0)^2$ & $(13 \AA/r_0)^2$ & $\xi_{coh}(\AA)$ & $(\xi_{coh}/r_0)^2$ &
$\xi_{coh}/r_0$   & $N_{ob}$ \\ \hline
    0.28  &    1.939         &     3.277        &      10          &   1.939
& 1.393 &  $\sim 1$ \\ \hline
    0.22      &  1.524       &     2.575        &      10         &    1.524
& 1.235 &  $\sim 2$ \\ \hline
    0.16      &  1.108       &     1.873        &      11         &    1.341
& 1.158 &  $\sim 3$ \\ \hline
  0.14      &    0.969       &     1.638        &      12         &    1.396
& 1.182 &  $\sim 3$ \\ \hline
  0.10      &    0.693       &     1.170        &      13         &    1.170
& 1.082 &  $\sim 6$ \\ \hline
  0.05      &    0.346       &     0.585       &       17         &    1.000
& 1.000 &  \\ \hline
  0.04      &    0.277       &     0.468       &       18         &    0.897
& 0.947 &  \\ \hline
  0.02      &    0.139       &     0.234       &       20         &    0.554
& 0.744 &  \\ \hline
\end{tabular}
\end{center}
\vskip .5cm \caption{The doping $x$ dependencies of NR charges. The
doping $x$ dependencies for $(10 \AA/r_0)^2$, $(13 \AA/r_0)^2$ at
fixed $\xi_{coh}=10 \AA$ and $\xi_{coh}=13 \AA$, respectively, for
the coherent length $\xi_{coh}$, the charge $(\xi_{coh}/r_0)^2$ and
the percolation parameter $\xi_{coh}/r_0$ at this $\xi_{coh}$ are
presented. The values for the number $N_{ob}$ of bosons surrounding
every fermion are shown in the last column.} \vskip -.5cm
\lab{tab-1}
\end{table}

As is seen in Table I, the charges $(10\AA/r_0)^2$, $(13\AA/r_0)^2$,
and $(\xi_{coh}/r_0)^2$ vary continuously with the doping $x$. This
is not surprising because they are functions of $r_0(x)$ and
$\xi_{coh}(x)$. From the analysis at the first critical doping,
$x_{c1}=0.05$, it follows that the charge $(\xi_{coh}/r_0)^2$ of the
visualized NR in Ref.~\ci{Gomes} equals $+1$. Thus, it
corresponds to the charge of a single hole. We note that at the critical
doping $x_{c1}=0.05$, the percolation parameter is given by
$\xi_{coh}/r_0=1.0$. This means the whole sample is entirely covered
with mini areas $\xi_{coh}^2=r_0^2$ contacting each other. It is
unexpected that at the second critical doping, $x_{c2}=0.28$, the
charge of the visualized NR takes the value $+2$. This implies that
at $\xi_{coh}^2=2r_0^2$ one has a pair of holes inside the NR and,
as a result, the superconductivity disappears completely. For
$x_{c2}=0.28$ we have $\xi_{coh}/r_0>1.0$, so that the charge
conductivity of the fermions still remains. As we presently understand,
the normal state charge conductivity is provided only by percolated 
fermions-holes and at temperature $T=0$ from $x_{c2}$ doping, 
while at $T\not=0$ by these fermions above the PG temperature boundary, 
for temperatures between $T_c$ and PG boundary there 
exists (fermion-boson) MIC.

Notice, that there are no particles in the nature with the
fractional charge, except the quasiparticles which can be produced
by many-body correlations like in the fractional quantum Hall
effect~\ci{Laughlin}. Hence, the problem having an
extra fractional charge present inside the NR has yet to be solved. We
are reminded in~\ci{Gomes,Pan} that PG-visualized NRs constitute the bulk
HTS phase below the critical temperature $T_c$, and
therefore, they are a precursor for that phase. This 
undoubtedly implies that the NRs represent bosons at least. At $x_{c1}=0.05$
one has the charge $(\xi_{coh}/r_0)^2=1$, so that one may surmise
that the NR represents just a boson localized in a square box
$\xi_{coh}^2$.

For $x>0.05$, the charge $(\xi_{coh}/r_0)^2$ has fractional part in addition to
that of the $+1$ (boson) part. We assign the former as the fractional part
of a fermion charge. Thus the total charge $(\xi_{coh}/r_0)^2$
of the NR includes a charge $+1$ for the boson and the fractional
charge for the fermion. However, as mentioned above, the
fractional charge can not exist. Therefore, we take the number
$N_{ob}$ of NRs to be equal to the inverse value of the fractional
part to form a charge of $+1$ for the fermion. As a result, we obtain
one fermion surrounding by $N_{ob}$ bosons. The values of $N_{ob}$
are outlined in the last column of the Table 1.

The NRs introduced in such a manner allow one to understand clearly the
fermion evolution over the following range $0.05 \leq x \leq 0.28$
of doping and to explain the origin of the second critical doping
$x_{c2}=0.28$. It is clear, that as $x$ increases, the number of fermions
increases inside the HTS phase. This means, that at $x_{c2}$
when the number of fermions becomes equal to the number of bosons,
one has a breaking of the boson percolation lines, and, thus the
HTS disappears.

\begin{figure}
\begin{center}
\includegraphics[width=9cm,scale=1]{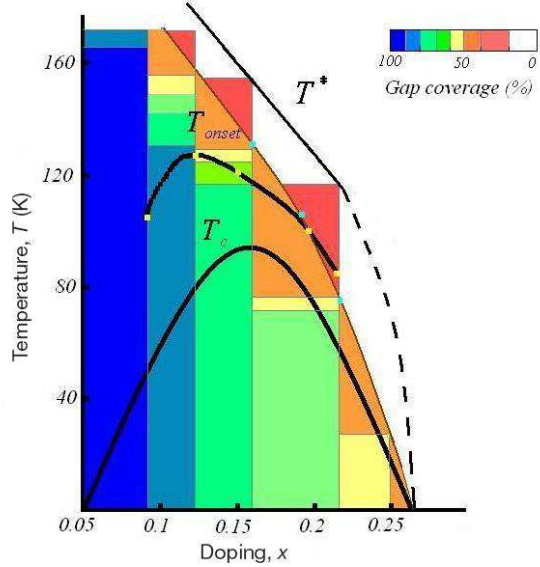}
\end{center}
\caption{Schematic single hole bosonic phase diagram for
$Bi_2Sr_2CaCu_2O_{8+\delta}$.} \lab{fig4}
\end{figure}

The schematic single hole bosonic phase diagram for
$Bi_2Sr_2CaCu_2O_{8+\delta}$ is depicted in the Fig. \ref{fig4}. The coloured
zones indicate the percentage of the sample that is gapped at given
temperature and doping. The solid lines correspond to the following
observed temperatures: PG boundary $T^*$ and onset temperature
$T_{onset}$ for Nernst effect signals taken from Ref.~\ci{Wang}, and
the critical temperature $T_c$ from Ref.~\ci{Gomes}. The
extrapolation for the connection of $T^*$ with the second critical
doping, $x_{c2}$, is depicted by the dashed line. The yellow points
correspond to fixed $T_{onset}$ values from Ref.~\ci{Wang}, and the
blue points represent the temperature data for $50\%$ of gapped area
of the sample from Ref.~\ci{Gomes} measured at fixed dopings. The
thin brown coloured solid line fits the blue points. The percentage
for the gapped doping is calculated by using the equation $(1-1/
(N_{ob}+1))\cdot 100\%$ under the assumption that the NRs overlap
each other. It is remarkable that $T_{onset}$ line is substantially
located in the brown coloured zones which means there is no bulk
bosonic property above these zones.

The important qualitative issue, which is a result of the experiment 
performed in~\ci{Gomes}, will now be discussed. The random positions in
real space of the observed pairs totally exclude any mechanism
for pair formation. Since they are occasionally positioned in this space
coherent excitations (phonons, magnons or other quasi-particles),
which create pairs, are problematic, if the system is homogenous.
The last observation deduced from the Gomes {\it et} {\it al.} paper is
the fundamental argument for the justification of the single hole
nature of the cuprate physics.

Summarizing the physics of the above sections, we are at the stage to 
formulate the main positions of our Coulomb single boson and single 
fermion two liquid model. 1. The doping charges, in the form of individual 
NRs, are embedded in the insulating parent compound of HTS copper oxides.
2. Before the first critical doping $x_{c1}$ with NR size $\xi_{coh}=17 \AA$, 
they are not percolated single bosons. 3. The origin of single bosons is in 
the anyon bosonization of 2D fermions. 4. At the first critical doping $x_{c1}$,
percolation of single boson NRs and thus HTS appears; there also appear from
$x_{c1}$ doping single fermions, but up to second critical doping $x_{c2}$ their
NRs do not percolate, thus single fermions between $x_{c1}$ and $x_{c2}$ are 
insulating. 5. The value $x_{c1}=0.05$ is universal for all copper oxide
HTSs, since percolating single boson NRs cover $50\%$ of a 2D sample area (like 
connecting squares in a chessboard); the same situation takes place with NRs for fermions 
at $x_{c2}$. 6.The normal phase charge conductivity appears from $x_{c2}$ at 
$T=0$ or above PG temperature boundary $T=T^*$, where the percolation of 
single fermions appears, while for temperatures between $T_c$ and $T^*$ there 
exists (fermion-boson) MIC. 7. The spatially rare charge density object, single 
boson, with NR size between $\xi_{coh}=17 \AA$ and $\xi_{coh}=10 \AA$, which 
correspond to $x_{c1}$ and $x_{c2}$ dopings, has zero total but fluctuating spin
inside the NR (this rareness also leads to a fluctuating charge inside the NR). 
8. Increasing the bosons' spin fluctuations with doping or 
temperature results in a transition of bosons into fermions, which occur at
PG $T^*$ or at $x_{c2}$. 9. At zero external magnetic field, the HTS is result 
of the Bose-Einstein condensate of single bosons. 10. At high external magnetic 
field, the PG insulating ground state is a result of the plasmon gas of these bosons.

\section{Hidden magnetic order and electronic nematicity experiments} 
\label{sec7}

First observed in $YBa_2Cu_3O_{6+x}$ (Y123) compound, using polarized elastic 
neutron diffraction~\ci{Fauque}, the  hidden magnetic order has been observed in 
three other copper oxide families: $HgBa_2CuO_{4+\delta}$ (Hg1201)~\ci{Li1,Li2}, 
$La_{2−x}Sr_xCuO_4$ (La214)~\ci{Baledent} and $Bi_2Sr_2CaCu_2O_{8+\delta}$ 
(Bi2212)~\ci{Almeida} (see for references also~\ci{Mangin}). Since a size of 
detection was a few structural cells, it was called the intra-unit-cell (IUC) 
hidden magnetic order.  

This interesting experiment has revealed the existence of IUC objects in the PG phase 
with fluctuating spin components inside, which exactly cancel each other,
so that the total spin of the every object was zero. Despite the authors of Ref.~\ci{Fauque}
having interpreted the physics of both, total and intra, spins either by invoking of a pair of 
oppositely flowing intra structural cell charge loop-currents or of staggered spins in 
the same cell, the role of these objects in the physics of copper oxides was not 
understood~\ci{Norman}.

This role becomes unambiguously clear, if we connect these objects with the visualization 
of NRs which exhibit the energy gap in the STM experiment~\ci{Gomes}. Since, as NRs, they
exist in the PG region and disappear in PG temperature boundary $T^*$~\ci{Fauque}. 
There is no doubt that their evolution with temperature will be the same as for NRs. 
While their evolution with doping (see Refs.~\ci{Fauque,Li1,Li2}) qualitatively 
coincides with that of NRs, described in~\ci{Abdullaev1}. However, minimal size NRs 
are single bosons, therefore, the IUC hidden magnetic order objects are also 
single bosons and PG and HTS pairs.  

On the other hand, the spatially intra rare charge density of each single boson 
allows one to understand the nature of the intra unit cell electronic nematicity 
(the dynamic charge fluctuations) which were recently observed in the STM experiment~\ci{Lawler}. The strong 
ferrielectric crystal field of the parrent compound forms an atomic scale charge distribution
within an individual NR. It is interesting that this distribution consists of fractional
charges, since the charge of NR is one, attached to each atom of parent cuprate, which can 
be seen in the experiment. 

The evolution of the dynamic charge fluctuations with doping resembles that of the IUC hidden magnetic 
order~\ci{Fujita} (Ref.~\ci{Fujita} contains a very thorough up-to-date list of references 
on the subject of IUC spin and charge fluctuations).

\section{MIC and insulating ground state} 
\label{sec8}

The experiemental investigation of lightly underdoped copper oxides, $La_{2-x}Sr_xCuO_4$,
$La_2CuO_4$, $YBa_2Cu_3O_x$, has revealed the MIC behavior in the $a - b$ plane resistivity
at zero magnetic field~\ci{Takagi}, when the resistivity decreased with lowering temperature 
and then began to increase from its minimum value. 

A similar MIC behavior for this resistivity has been observed for higher (up to critical) 
dopings in several classes of HTS cuprates~\ci{Ando}: $La_{2-x}Sr_xCuO_4$, electron doped 
$Pr_{2-x}Ce_xCuO_4$, $Bi_2Sr_{2-x}La_xCuO_{6+\delta}$, under strong magnetic fields (up to 60 T), 
when the HTS was suppressed. The temperature-MIC boundary in the temperature-doping phase diagram 
resembles the temperature-PG boundary dependence. 

The insulating ground state is intrinsic and robust, and therefore, belongs to class which is 
fundamental for understanding the whole of HTS cuprates-based physics. As we pointed out in 
the introduction, it may be a result of a gas of single bosons, more precisely, of a gas 
of 2D plasmons from single-charged bosons, which is insulating, when the Bose-Einstein condensate 
of these bosons vanishes under a strong magnetic field or at lowering and tending of dopimg
to the first critical value.

\section{Fermi pockets and stripe phases} 
\label{sec9}

Fermi pockets and stripe phases belong to a very relevant and highly controversial subject of the 
PG cuprate physics. The experimental situation is that Fermi pockets, except quantum 
oscillation experiments in a strong magnetic field~\ci{Sebastian}, have been detected 
in an ARPES experiment once in an underdoped PG region with doping $x=0.1$ and zero magnetic field 
above $T_c$ in $YBa_2Cu_3O_{6+x}$ (see Fig.2 of Ref.~\ci{Norman}). However, stripe 
phases (density waves) are not observed in all copper oxides~\ci{Vojta}.
Nevertheless, Fermi pockets and stripe phases are well established in cuprate-based physics 
and, probably, the comprehensive description of their state-of-art and critical analysis is 
made in the second review of Ref.~\ci{Vojta}. 

There is a widely accepted belief that there is a relationship between these two objects and 
furthermore, through Fermi surface reconstruction, stripe phases induce the formation of Fermi pockets 
(see reviews~\ci{Vojta,Norman} and references therein). We consider both these objects
as experimentally independent ones and try to understand their appearance within our 
Coulomb two liquid model.

Fermi pockets as constituents of a Fermi surface are a signature of fermion statistics. 
A manifestation of fermions in a parent compound's Brillouin zone starts at low dopings in the form
of Fermi arcs~\ci{Fujita}. Then, upon evolution with doping, it acquires Fermi pockets, 
and, finally, after optimal doping, it becomes circle like, as for a homogenous Fermi gas. 

The authors of quantum oscillation experiments, through the measurement of magnetoresistance 
oscillations at high magnetic fields in the underdome region of the temperature-doping phase diagram, 
have observed the Fermi liquid like behavior of conducting quasi-particles and that measured 
Fermi surface had a form of Fermi pockets. This effect was found for dopings up to optimal level
for the YBCO family of cuprates (see second review of Ref.~\ci{Vojta} for references, where there is
an indication that the quantum oscillations have also found in an electron doped cuprate). 
Unfortunately, this observation, under mentioned conditions, has not been reproduced by ARPES 
experiment. 

However, as we pointed out in Sec.~\ref{sec8}, for $x=0.1$ doping and temperatures between 
$T_c$ and PG boundary $T^*$ the system is close to the insulating state. Therefore, the Fermi pockets,
found by the ARPES experiment, belong to fermions in the insulating state. This finding confirms 
the position 4 of our Coulomb two liquid model (see end of Sec.~\ref{sec6}), i.e., that holes-fermions
between two critical dopings $x_{c1}$ and $x_{c2}$ are insulating, because of absence of 
percolation between their NRs. In our decsription two limiting values for the Fermi pocket wave
vector are approximately determined as $k_{F1}\approx 2\pi/\xi_{coh}(x_{c2})$ and  
$k_{F2}\approx 2\pi/\xi_{coh}(x)$, where $x$ is a current doping.  

As it was shown in Fig. 8 of Ref.~\ci{Caprara}, in which the schematic evolution of the 
nematic and stripe charge order phases with doping in underdoped cuprates is displayed, a nematic 
charge order, with violation of the system rotational symmetry and small spatial size of this order, 
dominates at initial dopings of the HTS dome phase diagram. As we discussed in Sec.~\ref{sec7}
the size of the nematic order coincides with the NR of single bosons. Therefore, the origin of 
this order is in the dynamical charge fluctuations inside the mentioned NR.

As is also seen from Fig. 8 of Ref.~\ci{Caprara}, stripe phase appears for dopings close to optimal
(critical) one, where NRs for sigle fermions dominate NRs for single bosons. This argument allows us 
to propose that the static charge order (stripe), which violates the parent lattice translational 
symmetry, is the result of insulating single fermions.  

\section{Why ground state of YBCO is Fermi liquid oscillating and of Bi-2212 is insulating?} 
\label{sec10}

As was mentioned in the previous section, the quantum oscillation experiment clearly indicates 
that the ground state of copper oxides from the YBCO family in a strong magnetic field is a 
Fermi liquid like, as is the case for conventional superconductors. Conversely, these copper oxides 
are absent in the list of materials (see section~\ref{sec8} or Ref.~\ci{Ando}), 
whose ground state is insulating in the same magnetic field. Does this mean that these quantum 
oscillating and insulating phenomena are the outcome of a single description for
cuprates? 

It seems, the answer to this question is "no" and it is in the following. 
The strong magnetic field "excites" a single boson from its ground state, so it 
occupies the first excited level of the system's fermion state. Physically, a NR with a single boson accupation
transits into NR with single fermion one. Therefore, all new formed single fermion NRs may provide 
percolation paths for the manifestation of the bulk Fermi liquid property, if the value for a critical doping
$x_c$ is low, as, for instance, the $x_c=0.19$ for YBCO HTS compounds. This is the reason why 
holes-fermions in the quantum oscillation experiment are conducting. However, originally, without 
a strong magnetic field, they existed in an insulating state.

However, if the value for a critical doping $x_c$ is large, as the $x_c=x_{c2}$ for 
$Bi-2212$, i.e., $Bi_2Sr_2CaCu_2O_{8+\delta}$ HTS compound, then single fermion NRs may not provide
percolation paths for a bulk Fermi liquid conductor and the ground state of $Bi-2212$ material in a 
strong magnetic field is insulating (see Ref.~\ci{Zavaritsky}).

\section{Non-Fermi liquid heat conductivity and entropy} 
\label{sec11}

The puzzling side of the PG normal phase is that in this region
some experimental data clearly demonstrate a failure of the Landau
Fermi liquid theory (LFLT), which is the basis of the theory of
normal metals. Hill $et$ $al.$~\ci{hill} reported that the
heat conductivity of the electron doped copper-oxide $Pr_{2-x} Ce_x
CuO_4$ measured at low temperature (low-T) deviated from the value
predicted by the LFLT, i.e., as the temperature decreases, the
temperature dependence of the heat conductivity ($\kappa$)
changes from a normal linear $\kappa \sim T$ behavior into an
anomalous $T^{3.6}$ one, which was described by the "downturn"
behavior of the heat conductivity. They also reported another
important non-Fermi liquid behavior: the Lorentz ratio of the
Wiedemann -- Franz law (WFL) in the region of the linear
T-dependence of $\kappa$ was significantly larger (1.7 times) than
Sommerfeld's value. These violations were also observed in the
$Bi_{2+x} Sr_{2-x} CuO_{6+\delta}$ copper-oxide in the vicinity of
the metal-insulator-crossover by Proust $et$ $al.$ \ci{proust}.

The normal state electronic specific heat, $c$, for superconductors $Y
Ba_2 Cu_3 O_{6+x}$ and $La_{2-x} Sr_x CuO_4$ above the high-$T_c$
transition temperature $T_c$ was experimentally investigated in
\ci{loram1} and \ci{loram2}, respectively.  Due to existence of
high-$T_c$ superconductivity, it is impossible to extract the
information on the low-$T$ dependence of the normal state $c$. On
the other hand, Loram $et$ $al.$ \ci{loram1} showed the
$T$-dependence of the entropy (${\cal S}$) ${\cal S} \sim T^i$ with
$i>1$ for the underdoped (insulating) material, which was driven
from the measured electronic specific heat, ignoring the superconducting
effects, while for the optimal doping compound ${\cal S} \sim T$ was
measured.

In Ref.~\ci{Abdullaev4} we tried to understand these non-Fermi
liquid properties of the low-$T$ heat conductivity, the specific
heat and entropy of copper-oxides within a Coulomb two-liquid model, 
consisting of single boson and fermion holes.

In Fig. \ref{fig5} we have plotted WFL and specific heat as a function of
temperature for compounds investigated experimentally in Refs.
\ci{hill} and \ci{loram1,loram2}. Experiment~\ci{hill} shows
$\kappa\sim T^{3.6}$ for normal state quasiparticles, while our
dependence is $\kappa\sim T^4$ and connected with the specific heat
dependence $c_1\sim T^4$ for the Coulomb Bose gas (single bosons).

\begin{figure}
\begin{center}
\includegraphics[angle=0,width=8.5cm,scale=1.0]{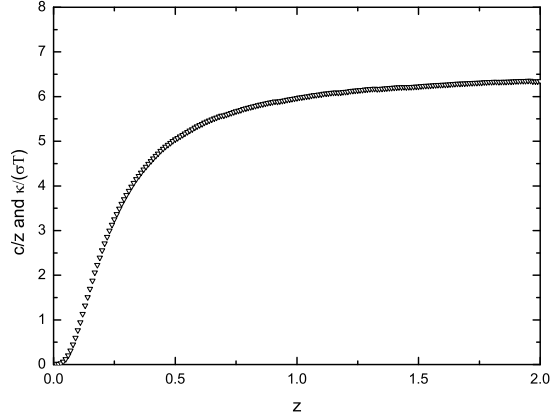}
\end{center}
\caption{The specific heat $c/z$ and WFL $\kappa/(\sigma T)$ as
function of $z$, where $z$ is linearly proportional to temperature.}
\lab{fig5}
\end{figure}

For the Lorentz ratio of the WFL we have derived in Ref.~\ci{Abdullaev4}
the formula \be L = L_0 \left[\dsf{3.106}{t^{1/3}}
\left(1-\dsf{t}{t_c} \right)+\dsf{t}{t_c}\right] \, , \lab{wf9} \ee
where $t$ and $t_c\approx0.19$ (see~\ci{tallon}) are dopings,
$L_0=(\pi^2/3) (k_B/e)^2$ is the Fermi liquid Sommerfeld's value of
the Lorentz ratio and the first term in square brackets originated
from the single boson WFL $\kappa/(\sigma T)= 3.106\cdot
L_0/t^{1/3}$. In the Fig. \ref{fig6} we displayed $L/L_0$ as function of $t$
in comparison with its values measured for different cuprates. The
good agreement for $L/L_0$ with experimental data is obvious.
\begin{figure}
\begin{center}
\includegraphics[angle=0,width=8.5cm,scale=1.0]{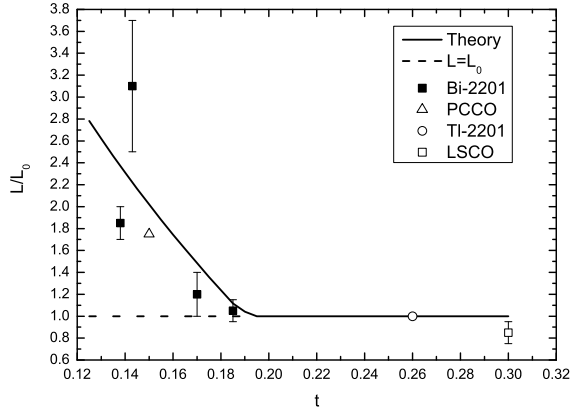}
\end{center}
\caption{ The Lorentz ratio $L/L_0$ (Eq.~\protect\re{wf9}) vs. $t$
(values for $t\ge 1$ are added artificially). Observed dots are from
Ref. \protect\ci{proust}.} \lab{fig6}
\end{figure}

In Ref.~\ci{Abdullaev4} we also obtained the expression of the
normal state entropy ${\cal S}$ 
\be {\cal
S}=\dsf{c_1}{4}\left(1-\dsf{T}{T^*}\right)+c_F \dsf{T}{T^*} \, ,
\lab{wf11} 
\ee 
where, $c_F$ is the heat capacity for a gas of fermions
and $T^*$ is the PG boundary temperature. The entropy, as function of
temperature at various dopings $t$, is depicted in Fig. \ref{fig7}. Comparing
the calculated dependencies for ${\cal S}$ with the experimental ones from
Ref.~\ci{loram1}, we see again the good agreement. As shown
in~\ci{Abdullaev4}, the nonlinear $T$ dependence of the entropy
${\cal S}$ at small temperatures is related to the insulating ground
state.

\begin{figure}
\begin{center}
\includegraphics[angle=0,width=8.5cm,scale=1.0]{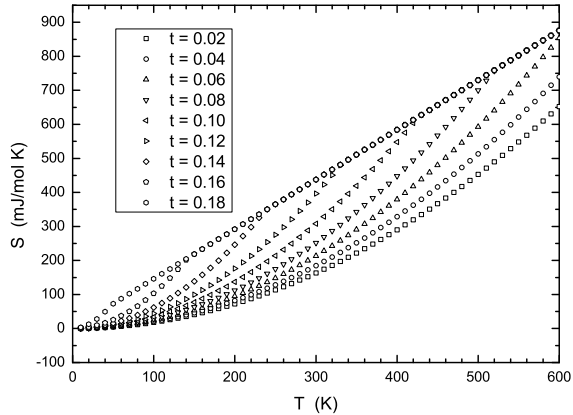}
\end{center}
\caption{ The entropy ${\cal S}$ (Eq.~\protect\re{wf11}) vs. $T$ at
various $t$ (values of ${\cal S}$ behind the crossing of linear and
nonlinear parts of ${\cal S}$ are added artificially).} \lab{fig7}
\end{figure}

\section{Conclusion}
\label{sec12}

In summary, we have formulated the Coulomb single
boson and single fermion two liquid model positions for HTS copper oxides: 
1. The doping charges, in the form of individual 
NRs, are embedded in the insulating parent compound of HTS copper oxides.
2. Before the first critical doping $x_{c1}$ with NR size $\xi_{coh}=17 \AA$, 
they are not percolated single bosons. 3. The origin of single bosons is in 
the anyon bosonization of 2D fermions. 4. At the first critical doping level $x_{c1}$,
the percolation of single boson NRs and thus HTS appears; there appear also from
$x_{c1}$ doping single fermions, but up to second critical doping level, $x_{c2}$, their
NRs do not percolate, thus single fermions between $x_{c1}$ and $x_{c2}$ are 
insulating. 5. The value $x_{c1}=0.05$ is universal for all copper oxide
HTSs, since percolating single boson NRs cover $50\%$ of the 2D sample area (like 
connecting squares in a chessboard); the same situation takes place with NRs for fermions 
at $x_{c2}$. 6. The normal phase charge conductivity appears from $x_{c2}$ at 
$T=0$ or above PG temperature boundary $T=T^*$, where the percolation of 
single fermions appears, while for temperatures between $T_c$ and $T^*$, there 
exists (fermion-boson) MIC. 7. The spatially rare charge density object, single 
boson, with NR size between $\xi_{coh}=17 \AA$ and $\xi_{coh}=10 \AA$, which 
correspond to $x_{c1}$ and $x_{c2}$ dopings, has zero total but fluctuating 
inside of NR spin (this rareness leads also to fluctuating charge inside of NR). 
8. The increase of boson spin fluctuations with doping or 
temperature results to a transition of bosons into fermions, which occurs at
PG $T^*$ or at $x_{c2}$. 9. At zero external magnetic field, the HTS is a result 
of the Bose-Einstein condensate of single bosons. 10. At high external magnetic 
field, the PG insulating ground state is the result of a plasmon gas consisting of these bosons.

Within these positions, we have succeeded in understanding the
following constituents of the doping-temperature phase diagram for
 hole doped copper oxides: (i) the first and second critical
doping levels are a result of the emergence and disappearance of the
single hole boson percolation lines, respectively; (ii) the
disappearance of the percolation lines leads to the end of the PG
bulk bosonic property or to the end of Nernst effect signals; (iii)
the fact that the PG boundary was bound, where the single hole
bosons disappear, was confirmed by Ref.~\ci{Gomes}. Our findings are
consistent with the recent observation~\ci{Gavrilkin} of the
superconducting phase which consisted of an array of nanoclusters
embedded in an insulating matrix and of the percolative transition to
this phase from the normal phase in $YBa_2Cu_3O_{6+ \delta}$.
A recent experiment~\ci{Zeng} also displayed the percolative
superconductor-insulator quantum phase transition in the 
electron-doped $Pr_{2-x}Ce_xCuO_4$. 
Superconducting islands, introduced in insulating background, have
been used to interpret the superconductor-insulator
transition in $Bi_2Sr_{2-x}La_xCaCu_2O_{8+\delta}$ compound~\ci{Oh}.

We are reminded of the possible scenario for the origin of PG phase in the
copper-oxides. In Ref.~\ci{Abdullaev2}, we pointed out that the PG
boundary exactly coincides in the experiment with the structural
phase transition line, where the symmetry of the sample structure
changes. However, the structural phase transition induces a mechanical 
strain in the system. This mechanical strain changes the
magnetic phase transition, existing in a system, from a second
order into a first order one. However, this first order phase
transition is close to the second order one. This effect referred to in
the literature as a striction. The phase transition of single bosons
into fermions, discussed in Sec.~\ref{sec4}, is possibly governed
by a striction. Therefore, transition peaks in the specific heat
increment are washed out in the cuprate underdoping regime~\ci{loram1}. 
On the other hand, the first order phase
transition accepts the existence of a meta-stable phase. We
believe that PG phase of cuprates is a meta-stable phase of single
bosons, which effective spins are fluctuating and interacting with
each others. At the PG boundary this interaction entirely destroys
bosons, transforming them into fermions.

We predict the existence of non-percolated single bosons
at low-$T$ before the first critical doping $x_{c1}$ in the
doping-temperature phase diagram of copper-oxides, which is already 
seen in the experiment (see below). For these
dopings, a scanning tunneling microscopy measurement may probe the
same picture for minimal size NRs as for PG region, close to PG
boundary. The Bose statistics of these particles may be
experimentally detected by some methods described in
Ref.~\ci{Stern1}. These methods might be also applied to detect the
insulating fermions inside HTS dome, which is the main hypotheses of the
present treatment. 

We predict the existence of quantum oscillations in the ground state 
for copper oxides in a strong magnetic field with low values of the critical 
doping $x_c$ and an insulating phase for HTS compounds having large $x_c$ values.

Our Coulomb two liquid model also predicts there to be no Josephson and Andreev effects 
between the two $a - b$ planes of underdoped HTS, belonging to two cuprates and separated by 
an insulating layer. It would be interesting to probe the atomic scale fractional
charges in the STM experiment for electronic nematicity.

Recently, authors of Ref.~\ci{Mclaughlin} have reported on the emergent transition 
for superconducting fluctuations in the deep antiferromagnetic phase at a 
remarkably low critical doping level, $x_c= 0.0084$, for ruthenocuprates, 
$RuSr_2(R,Ce)_2Cu_2O_{10-\delta}$ with $R = Gd,Sm,Nd$. In this paper, it was claimed 
that those fluctuations have an intrinsic electronically-inhomogenous nature and 
provide new support for bosonic models of the superconducting mechanism.

At the end, we point out that the present paper is a substantially extended 
and critically revised version of our previous Ref.~\ci{Abdullaev6}, in which
we added the alternative physical insight into the background of HTS physics.

\section*{Acknowledgements}

Authors B. Abdullaev and C.-H. Park acknowledge the support of the research by the National Research 
Foundation (NRF) Grant (NRF-2013R1A1A2065742) of the Basic Science Research Program 
of Korea.

\end{document}